\documentclass[conference]{IEEEtran}

\pdfoutput=1

\usepackage{graphicx}
\usepackage{epstopdf}
\usepackage{amssymb}
\usepackage[cmex10]{amsmath}
\usepackage{amsmath}
\usepackage[font=footnotesize]{subfig}
\usepackage{mathtools}
\usepackage{tikz}
\usetikzlibrary{shapes}
\usepackage{pgfplots}
\tikzset{>=latex}
\usepackage{setspace}
\usepackage{afterpage}

\usepackage{multicol}

\usepackage{cite}

\usepackage{caption}

\def\BibTeX{{\rm B\kern-.05em{\sc i\kern-.025em b}\kern-.08em T\kern-.1667em\lower.7ex\hbox{E}\kern-.125emX}}

\usetikzlibrary{patterns}

\newcounter{tempEquationCounter} 
\newcounter{thisEquationNumber}

\begin{document}
	
\bstctlcite{IEEEexample:BSTcontrol} 

\title{Age of Information and Throughput in a Shared Access Network with Heterogeneous Traffic}

\author{
	\IEEEauthorblockN{Antzela~Kosta\IEEEauthorrefmark{1}, Nikolaos~Pappas\IEEEauthorrefmark{1}, Anthony~Ephremides\IEEEauthorrefmark{1}\IEEEauthorrefmark{2}, and~Vangelis~Angelakis\IEEEauthorrefmark{1}}
	\IEEEauthorblockA{\IEEEauthorrefmark{1} Department of Science and Technology, Link{\"o}ping University, Campus  Norrk{\"o}ping, 
		60 174, Sweden}
	\IEEEauthorblockA{\IEEEauthorrefmark{2} Electrical and Computer Engineering Department, University of Maryland, College Park, MD 20742\\
		E-mail: \{antzela.kosta, nikolaos.pappas, vangelis.angelakis\}@liu.se,  etony@umd.edu}  }

\maketitle

\begin{abstract}
We consider a cognitive shared access scheme consisting of a high priority primary node and a low priority network with $N$ secondary nodes accessing the spectrum. 
Assuming bursty traffic at the primary node, saturated queues at the secondary nodes, and multipacket reception capabilities at the receivers, we derive analytical expressions of the time average age of information of the primary node and the throughput of the secondary nodes. We formulate two optimization problems, the first aiming to minimize the time average age of information of the primary node subject to an aggregate secondary throughput requirement. The second problem aims to maximize the aggregate secondary throughput of the network subject to a maximum time average staleness constraint.
Our results provide guidelines for the design of a multiple access system with multipacket reception capabilities that fulfills both timeliness and throughput requirements.

\end{abstract}
\IEEEpeerreviewmaketitle
\section{Introduction}

A fundamental problem of multiaccess communications is how to efficiently share the channel resource among multiple devices.
In cognitive radio terminology, primary users are defined as the users who have higher priority on the access of a specific part of the spectrum. On the other hand, secondary users have lower priority and access the medium through spectrum sensing that detects bands that are not occupied by the primary users. In reality, interference caused by the secondary users is unavoidable and it should be included in a more realistic model.

In this work, we consider a cognitive network with multipacket reception (MPR) capabilities, where the secondary nodes make transmission attempts with a given probability.
With random access the uncoordinated secondary nodes can transmit simultaneously with the primary node allowing multiple successful receptions.
The access probabilities should be appropriately chosen to mitigate the impact on the performance of the primary node.
The literature so far considers delay or throughput as the performance metric of the primary node, however a concept that captures timeless more precisely and has not been considered yet in the cognitive shared access context in \emph{age of information} (AoI). 

The concept of AoI was introduced in \cite{Kaul12_INFOCOM} to quantify the freshness of the knowledge a monitor has about the status of a remote system.
Consider a source-destination communication pair.
With AoI the freshness is quantified, at any moment, as the time that elapsed since the last received status update was generated by the source \cite{NET-060}.
Interestingly, timely updating a destination about a remote system is neither the same as maximizing the utilization of the communication system, nor of ensuring that generated status updates are received with minimum delay \cite{Bedewy16_ISIT}.
Moreover, it has been proven in \cite{Yates15_ISIT} and \cite{Sun16_INFOCOM} that transmitting a status update as soon as a preceding update finishes service is not necessarily optimal with respect to the average AoI of the system.
Such a policy achieves the maximum throughput and the minimum delay but is not optimal with respect to AoI.

Among the works that studied AoI in scheduling \cite{Kadota2018_INFOCOM,Kadota2016_Allerton,Hsu17_ISIT,Joo17_WiOpt}, the authors in \cite{Kaul11_SECON} considered the minimization of age of status updates sent by vehicles over a carrier-sense multiple access (CSMA) network.
In \cite{Kaul17_ISIT}, scheduled access and slotted ALOHA-like random access with respect to AoI is considered, however the queueing aspect along with random access is not captured.
In this work, we assume that the primary node in the multiple access system has unlimited buffer capacity to store newly generated and backlogged packets.
The associated receiver is interested in timely status updates from the primary node, i.e., minimum AoI or keeping the AoI below a threshold.
In addition to providing good performance to the primary node, we impose a
minimum aggregate throughput guarantee for the low priority network.
Throughput and delay performance in cognitive shared access networks with queueing analysis has been studied in \cite{Chen18_transactions,Pappas2014throughput}.

\subsection{Contribution}
We focus on the joint throughput-timeliness performance of a cognitive shared access network consisting of a primary link and a large secondary network.
Assuming bursty traffic at the primary node, saturated queues at the secondary nodes, and multipacket reception capabilities at the receivers, we derive analytical expressions of the time average age of information of the primary node and the throughput of the secondary nodes as functions of the primary and secondary access probabilities.
We introduce the optimization problem of minimizing the time average AoI of the primary node subject to an aggregate secondary throughput requirement.
Moreover, a second problem aims to maximize the aggregate secondary throughput of the network subject to a maximum time average staleness constraint.
The numerical evaluation of the problems shows how the optimal access probabilities and the resulting network objectives vary with the system parameters and highlights the characteristics of the AoI as a performance index.

\section{System Model}
\label{sec:model}

\subsection{Network Model}
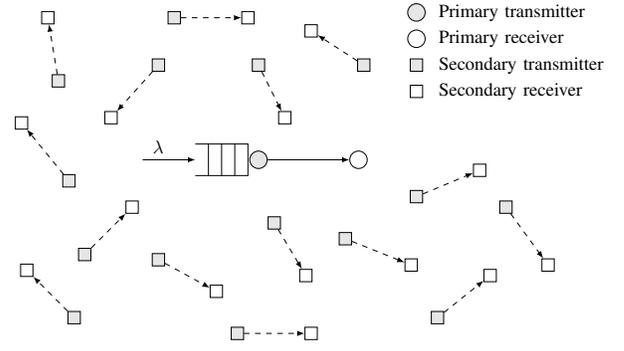
\begin{figure}[t!]
	\centering
	\scalebox{.7}{\begin{tikzpicture}


 \draw[]  (2.8,4) -- (3.8,4) -- (3.8,3.4) -- (2.8,3.4) ;
 \draw[]  (3.3,4) -- (3.3,3.4) ;
 \draw[]  (3.55,4) -- (3.55,3.4) ;
 \draw[]  (3.05,4) -- (3.05,3.4) ;

\node at (4.0,3.7) [circle,draw,fill=gray!20] (TxPN) {};
\node at (5.9,3.7) [circle,draw] (RxPN) {};

\draw[->]  (1.8,3.7) -- (2.8,3.7) node[pos=.3,sloped,above] {$\lambda$};
\draw[->]  (TxPN) -- (RxPN);


\node at (0.7,1.9) [rectangle,draw,fill=gray!20] (TxSN1) {};
\node at (1.6,2.8) [rectangle,draw] (RxSN1) {};

\node at (2.1,1.8) [rectangle,draw,fill=gray!20] (TxSN2) {};
\node at (3.2,1.2) [rectangle,draw] (RxSN2) {};

\node at (7,3) [rectangle,draw,fill=gray!20] (TxSN3) {};
\node at (8.2,3.5) [rectangle,draw] (RxSN3) {};

\node at (6,5.5) [rectangle,draw,fill=gray!20] (TxSN4) {};
\node at (5,6.15) [rectangle,draw] (RxSN4) {};

\node at (7.4,0.7) [rectangle,draw,fill=gray!20] (TxSN5) {};
\node at (8.4,1.5) [rectangle,draw] (RxSN5) {};

\node at (2.1,5.5) [rectangle,draw,fill=gray!20] (TxSN6) {};
\node at (1.2,4.5) [rectangle,draw] (RxSN6) {};

\node at (4.0,5.5) [rectangle,draw,fill=gray!20] (TxSN7) {};
\node at (4.5,4.5) [rectangle,draw] (RxSN7) {};

\node at (4.3,2.5) [rectangle,draw,fill=gray!20] (TxSN8) {};
\node at (4.9,1.5) [rectangle,draw] (RxSN8) {};



\node at (8.7,2.8) [rectangle,draw,fill=gray!20] (TxSN9) {};
\node at (9.5,1.7) [rectangle,draw] (RxSN9) {};

\node at (5.65,2.2) [rectangle,draw,fill=gray!20] (TxSN10) {};
\node at (6.9,1.7) [rectangle,draw] (RxSN10) {};

\node at (3.6,0.4) [rectangle,draw,fill=gray!20] (TxSN11) {};
\node at (5.0,0.4) [rectangle,draw] (RxSN11) {};

\node at (0.4,3.3) [rectangle,draw,fill=gray!20] (TxSN12) {};
\node at (-0.5,4.4) [rectangle,draw] (RxSN12) {};

\node at (2.4,6.4) [rectangle,draw,fill=gray!20] (TxSN13) {};
\node at (3.8,6.4) [rectangle,draw] (RxSN13) {};

\node at (0.5,0.7) [rectangle,draw,fill=gray!20] (TxSN14) {};
\node at (-0.4,1.6) [rectangle,draw] (RxSN14) {};



\node at (0.2,5.2) [rectangle,draw,fill=gray!20] (TxSN15) {};
\node at (0.00,6.4) [rectangle,draw] (RxSN15) {};

\node at (7,6.5) [circle,draw,fill=gray!20] (TxPNlabel) {};
\node at (7,6.0) [circle,draw] (RxPNlabel) {};
\node at (7,5.5) [rectangle,draw,fill=gray!20] (TxSNlabel) {};
\node at (7,5.0) [rectangle,draw] (RxSNlabel) {};

\draw[]  (TxPNlabel) node[right=0.3cm] {Primary transmitter};
\draw[]  (RxPNlabel) node[right=0.3cm] {Primary receiver};
\draw[]  (TxSNlabel) node[right=0.3cm] {Secondary transmitter};
\draw[]  (RxSNlabel) node[right=0.3cm] {Secondary receiver};

\draw[->,dashed]  (TxSN1) -- (RxSN1) ;
\draw[->,dashed]  (TxSN2) -- (RxSN2) ;
\draw[->,dashed]  (TxSN3) -- (RxSN3) ;
\draw[->,dashed]  (TxSN4) -- (RxSN4) ;
\draw[->,dashed]  (TxSN5) -- (RxSN5) ;
\draw[->,dashed]  (TxSN6) -- (RxSN6) ;
\draw[->,dashed]  (TxSN7) -- (RxSN7) ;
\draw[->,dashed]  (TxSN8) -- (RxSN8) ;
\draw[->,dashed]  (TxSN9) -- (RxSN9) ;
\draw[->,dashed]  (TxSN10) -- (RxSN10) ;
\draw[->,dashed]  (TxSN11) -- (RxSN11) ;
\draw[->,dashed]  (TxSN12) -- (RxSN12) ;
\draw[->,dashed]  (TxSN13) -- (RxSN13) ;
\draw[->,dashed]  (TxSN14) -- (RxSN14) ;
\draw[->,dashed]  (TxSN15) -- (RxSN15) ;



\end{tikzpicture}}
	\caption{A cognitive network with a primary source-destination pair and a set of secondary pairs.}
	\label{fig:network}
\end{figure}

We consider a cognitive shared access network consisting of one primary 
source-destination pair and $N$ secondary pairs all located at fixed positions over the network, as shown in Fig.~\ref{fig:network}.
We assume packet based communication.
The primary node has a buffer of infinite capacity to store incoming packets.
Packets have equal length and time is divided into slots such that the transmission time of a packet is equal to one slot. 
Each packet arriving at the primary destination provides a status update of the primary source.
The primary node generates status updates according to a Bernoulli process with average arrival rate $\lambda$.
The status update generation process is assumed to be independent over slots.
The queues at the secondary nodes are assumed to be saturated, that is, there is always a packet waiting for transmission.

Provided that the buffer is not empty, at the beginning of each time slot, the primary node attempts to transmit the packet at the head of the queue with probability $q_{pr}$.
In addition, the secondary nodes attempt to access the spectrum/channel with probability $q_s$, for all nodes. 
We consider a generalization of the collision channel, where the receivers have multipacket reception (MPR) capabilities and the secondary nodes can transmit simultaneously with the primary node using non-orthogonal spectrum resources.
Acknowledgments of successful transmissions are assumed instantaneous and error-free.
In case of failure, the node stores the missed packet into its queue and attempts retransmission.

\subsection{Physical layer model}
In a multiple access scheme with MPR capabilities the nodes might interfere with each other. 
Then, a successful reception requires that the signal-to-interference-plus-noise ratio at the $i$th receiver, $SINR_i$, exceeds a certain threshold $\gamma_d$.
The wireless channel is modeled as Rayleigh fading channel with additive white Gaussian noise.

Let $P_{tx}(i,d)$ be the transmit power of node $i$ and $r(i,d)$ be the distance between $i$ and the destination $d$.
Then, the power received by $d$ when $i$ transmits is $P_{rx}(i,d) = A(i,d) h(i,d)$, where $A(i,d)$ is a unit-mean exponentially distributed random variable  representing channel fading.
The received power factor $h(i,d)$ is given by $h(i,d) = P_{tx}(i,d) (r(i,d))^{-\alpha}$ where $\alpha$ is the path loss exponent.
The success probability of link $i-d$ when the transmitting nodes are in $\mathcal{K}$ is given by 
\begin{align}
P^d_{i/\mathcal{K}} = & \exp \left( -\frac{\gamma_d \eta_d}{v(i,d) h(i,d)} \right) \notag \\
& \prod_{k \in \mathcal{K}\backslash \{i,d\}} \left( 1+ \gamma_d \frac{v(k,d) h(k,d)}{v(i,d) h(i,d)} \right)^{-1},
\label{eq:success_Pr}
\end{align}
where $v(i,d)$ is the parameter of the Rayleigh fading random variable, and $\eta_d$ denotes the received noise power at $d$.
Equation \eqref{eq:success_Pr} can be reformulated for the case of a symmetric $N$-nodes network. 
Then, the probability of success of link $i-d$ when $k$ nodes transmit is
\begin{equation}
P^d_{i/k}  =  \exp \left( -\frac{\gamma_d \eta_d}{v(i,d) h(i,d)} \right) \left( \frac{1}{1 + \gamma_d} \right)^{k-1}.
\label{eq:symmetric_success_Pr}
\end{equation}

\subsection{Considered metrics}

We consider two performance metrics.
The secondary nodes are interested in maximizing their throughput.
The maximum throughput of a secondary link is the maximum service rate $\mu_s$ in packets/slot that can be achieved over the link since the secondary nodes are assumed to be saturated.
The primary node is interested in the freshness of the knowledge the destination has about its status. This freshness can be captured by the concept of the AoI of the primary node.
In the next sections, we formally characterize both performance metrics and derive their analytical expressions as functions of the access probabilities $q_{pr}$ and $q_s$. 

\section{Age of Information formulation}

\afterpage{%
	\begin{figure}[h!]
		\centering
		\begin{tikzpicture}[scale=0.90]
\draw[->] (0,0) -- (8.2,0) node[anchor=north] {$t$};
\draw[->] (0,0) -- (0,3.5) node[anchor=east] {$\Delta_t$};

\draw	(-0.3,0.38) node[anchor=south] {$\Delta_0$};

 

\draw[fill=gray!20] (0,0.0) -- (0,1.0) -- (0.5,1)-- (0.5,1.5)-- (0.5,1.5)-- (1.0,1.5)-- (1.0,0.5) -- (0.5,0.5) -- (0.5,0.0);
 
\draw[fill=gray!20]  (1.5,0) -- (1.5,0.5) -- (2.0,0.5) -- (2.0,1)-- (2.5,1) -- (2.5,1.5)-- (3.0,1.5)-- (3.0,2)-- (3.5,2.0)-- (3.5,2.5)-- (4.0,2.5)-- (4.0,3.0)-- (4.5,3.0)-- (4.5,2.5)-- (4.0,2.5)-- (4.0,2.0)-- (3.5,2)-- (3.5,1.5)-- (3.0,1.5)-- (3.0,1) -- (2.5,1)-- (2.5,0.5)-- (2,0.5)-- (2,0);

\draw[pattern=crosshatch dots, pattern color=gray!20] (0.5,0)-- (0.5,0.5)-- (1,0.5)--   (1.0,1)-- (1.5,1) -- (1.5,1.5)-- (2.0,1.5)-- (2.0,2)-- (2.5,2.0)-- (2.5,1.0)-- (2.0,1.0)-- (2.0,0.5)-- (1.5,0.5)-- (1.5,0.0);

\draw[pattern=crosshatch dots, pattern color=gray!20]  (2.0,0.0) -- (2.0,0.5) -- (2.5,0.5) -- (2.5,1.0) -- (3.0,1.0) -- (3.0,1.5) -- (3.5,1.5) -- (3.5,2.0) -- (4.0,2.0) -- (4.0,2.5) -- (4.5,2.5) -- (4.5,3.0) -- (5.0,3.0) -- (5.0,3.5)  -- (5.5,3.5) -- (5.5,1.5) -- (5.0,1.5) -- (5.0,1.0) -- (4.5,1.0) -- (4.5,0.5) -- (4.0,0.5)  -- (4.0,0.0);

\draw[white] (5.5,3.5) -- (5.5,1.5) ;

\draw[fill=gray!20] (6.0,0.0) -- (6.0,0.5) --  (6.5,0.5) -- (6.5,1)-- (7.0,1) -- (7.0,1.5)-- (7.5,1.5)-- (7.5,0.5) -- (7.0,0.5) -- (7.0,0.0);

\draw	(-0.5,0) node[anchor=north] {$t_0$}
           (0.5,0) node[anchor=north] {$t_1$}
		    (1.5,0) node[anchor=north] {$t_2$}
		    (2,0) node[anchor=north] {$t_3$}
		    (4.0,0) node[anchor=north] {$t_4$}
		    (6,0) node[anchor=north] {$t_{n-1}$}
		    (7.0,0) node[anchor=north] {$t_n$};
		    
\draw[->,>=stealth]    (1,0) -- (1,-0.4) node[anchor=south,below] {$t'_1$};
\draw[->,>=stealth]  (2.5,0) -- (2.5,-0.4) node[anchor=south,below] {$t'_2$};
\draw[->,>=stealth]  (4.5,0) -- (4.5,-0.4) node[anchor=south,below] {$t'_3$};
\draw[->,>=stealth]   (7.5,0) -- (7.5,-0.4) node[anchor=south,below] {$t'_n$};
		    	    		 
\draw	(0.55,0.75) node{{\scriptsize $J_1$}}
		    (1.5,0.75) node{{\scriptsize $J_2$}};
\draw   (3.5,0.75) node{{\scriptsize $J_4$}}
           (7,0.75) node{{\scriptsize $J_n$}};
 \draw   (7.25,0.25) node{{\scriptsize $\tilde{J}$}};
           
\draw[<-] (3.3,1.6) to [out=95,in=250] (3.3,2.5) node [above] {{\scriptsize $J_3$}};           
           
\draw [thick](0.5,-1.2) -- (1.5,-1.2) node[pos=.5,sloped,below] {$Y_2$} ;
\draw[thick]  (0.5,-1.3) -- (0.5,-1.1); 
\draw [thick](1.5,-1.2) -- (2.5,-1.2) node[pos=.5,sloped,below] {$T_2$} ;
\draw[thick]  (1.5,-1.3) -- (1.5,-1.1) 
                    (2.5,-1.3) -- (2.5,-1.1);
                    
\draw [thick](6,-1.2) -- (7.0,-1.2) node[pos=.5,sloped,below] {$Y_n$} ;
\draw[thick]  (6,-1.3) -- (6,-1.1); 
\draw [thick](7.0,-1.2) -- (7.5,-1.2) node[pos=.5,sloped,below] {$T_n$} ;
\draw[thick]  (7.0,-1.3) -- (7.0,-1.1) 
                    (7.5,-1.3) -- (7.5,-1.1);
 
\draw[thick] (0,0.5) -- (0,1.0) -- (0.5,1)-- (0.5,1.5)-- (0.5,1.5)-- (1.0,1.5)-- (1.0,1.0);

\draw[thick] (1.0,1)-- (1.5,1) -- (1.5,1.5)-- (2.0,1.5)-- (2.0,2)-- (2.5,2.0)-- (2.5,1.5);
\draw[thick] (2.5,1) -- (2.5,1.5)-- (3.0,1.5)-- (3.0,2)-- (3.5,2.0)-- (3.5,2.5)-- (4.0,2.5)-- (4.0,3.0)-- (4.5,3.0);
\draw[thick]  (4.5,2.5)-- (4.5,3.0)-- (5.0,3.0)-- (5.0,3.5)-- (5.5,3.5);

\draw[thick] (6.5,0.5) -- (6.5,1)-- (7.0,1) -- (7.0,1.5)-- (7.5,1.5)-- (7.5,0.5);

\draw[dotted] (1,0) -- (1,1.5);
\draw[dotted] (2.5,0) -- (2.5,2.0); 
\draw[dotted] (4.5,0) -- (4.5,3.0); 
\draw[dotted] (7.5,0) -- (7.5,0.5); 
\draw[dotted] (-0.5,0) -- (-0.5,0.5)-- (0,0.5); 
\draw[dotted] (0.5,0) -- (0.5,0.5) -- (1.0,0.5) -- (1.0,1); 
\draw[dotted] (1.5,0) -- (1.5,0.5) -- (2.0,0.5) -- (2.0,1)-- (2.5,1); 
\draw[dotted] (2,0) -- (2,0.5) -- (2.5,0.5) -- (2.5,1)-- (3.0,1) -- (3.0,1.5)-- (3.5,1.5)-- (3.5,2)-- (4.0,2.0)-- (4.0,2.5); 

\draw[dotted] (4,0) -- (4,0.5) -- (4.5,0.5)-- (4.5,1.0)-- (5.0,1.0); 
\draw[dotted] (6,0) -- (6,0.5) -- (6.5,0.5); 
\draw[dotted] (7.0,0) -- (7.0,0.5) -- (7.5,0.5); 

\draw[thick]  (5.2,-0.15) -- (5.2,0.15) 
                    (5.3,-0.15) -- (5.3,0.15);
\draw[white, fill=white!50] (5.21,-0.2) -- (5.21,0.2) -- (5.29,0.2) -- (5.29,-0.2) ;                   

\end{tikzpicture}
		\caption{Example of AoI evolution of the primary node at the receiver.}
		\label{fig:age_vs_time_slotted}
\end{figure}
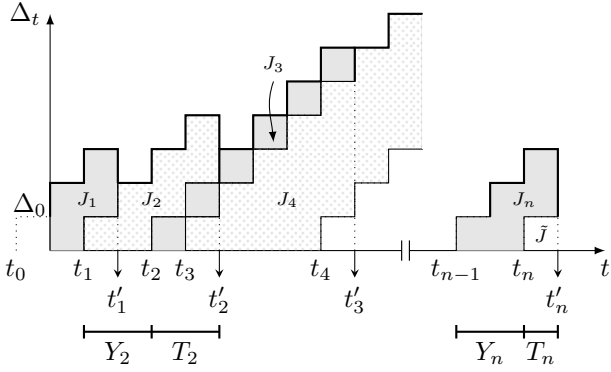}

Assume status update $j$ of the primary node is generated at time $t_{j}$ and is received by the destination at time $t_{j}^{'}$.
Then, $T_{j} = t'_{j}  - t_{j}$ is the system time of update $j$ of the primary node, corresponding to the sum of the queue waiting time and the service time.
The $j$th interarrival time of the primary node is defined as the
random variable $Y_{j} = t_{j} - t_{{j}-1}$.

The AoI of the primary node at the receiver is defined as the random process $\Delta_t = t-u(t)$, where $u(t)$ is the timestamp of the most recently received update.
Fig.~\ref{fig:age_vs_time_slotted} shows an illustrative example of the evolution of AoI in time.
Without loss of generality, assume that at $t=0$ we start observing the system, the queues are empty, and the AoI of the primary node at the destination is $\Delta_0$.
Between $t_{j-1}^{'}$ and $t_{j}^{'}$, where there is an absence of updates from the primary node, the AoI increases linearly with time. 
Upon reception of a status update from the primary node the AoI is reset to the delay that the packet experienced going through the transmission system. 

Ensuring the average AoI of the primary node is small corresponds to maintaining information about the status of the node at the destination fresh.
Given an age process $\Delta_t$ and assuming ergodicity, the average age can be calculated using a sample average that converges to its corresponding stochastic average.
For an interval of observation $(0,\mathcal{T})$, the time average age of the primary node is
	\begin{equation}
	\Delta_{\mathcal{T}}=\frac{1}{\mathcal{T}}\sum_{t=0}^{N(\mathcal{T})}\Delta_t,
	\label{eq:average_age}
	\end{equation}
when we assume that the observation interval ends with the service completion of $N(\mathcal{T})$ packets.
The summation in \eqref{eq:average_age} can be calculated as the area under $\Delta_t$.
Then, the time average age can be rewritten as a sum of disjoint geometric parts.
Starting from $t=0$, the area is decomposed into the area $J_1$, the areas $J_j$ for $j=2, 3, \ldots N(\mathcal{T})$, and the area of width $T_n$ that we denote $\tilde{J}$.
Then, the decomposition of $\Delta_{\mathcal{T}}$ yields
\begin{align}
\Delta_{\mathcal{T}} = & \frac{1}{\mathcal{T}} \left( J_1 + \tilde{J} + \sum_{j=2}^{N(\mathcal{T})} J_j \right) \notag \\
= &  \frac{J_1 + \tilde{J}}{\mathcal{T}} + \frac{N(\mathcal{T})-1}{\mathcal{T}} \frac{1}{N(\mathcal{T})-1} \sum_{j=2}^{N(\mathcal{T})} J_j.
\label{eq:Delta_T}
\end{align}
The time average $\Delta_{\mathcal{T}}$ tends to the ensemble \textit{average age} as $\mathcal{T} \rightarrow\infty$, i.e.,
\begin{equation}
\Delta = \lim_{\mathcal{T} \rightarrow\infty}\Delta_{\mathcal{T}}\footnote{We assume that the existence of the limit is guaranteed by the stability conditions discussed in the next section.}.
\label{eq:Delta_avglimit}
\end{equation}
Note that the term $(J_1 + \tilde{J})/\mathcal{T}$ goes to zero as $\mathcal{T}$ grows and also let
\begin{equation}
\lambda = \lim_{\mathcal{T} \rightarrow\infty} \frac{N(\mathcal{T})}{\mathcal{T}}
\label{eq:lambda}
\end{equation}
be the steady state rate of status updates generation.
Furthermore, using the definitions of the interarrival and system times, we can write the areas $J_j$ as
\begin{align}
J_j & = \sum_{m=1}^{Y_j+T_j} m - \sum_{m=1}^{T_j} m  \notag \\ 
& = \frac{1}{2} (Y_j+T_j)(Y_j+T_j+1) - \frac{1}{2} T_j(T_j+1)  \notag \\
& = Y_j T_j + Y_j^2/2 + Y_j/2.
\label{eq:J_i}
\end{align}
Then, substituting \eqref{eq:Delta_T}, \eqref{eq:lambda}, and \eqref{eq:J_i}, to \eqref{eq:Delta_avglimit} the average age of information of the primary node is given by
\begin{equation}
\Delta = \lambda\: \left( \mathbb{E}[YT]+\frac{\mathbb{E}[Y^2]}{2}+ \frac{\mathbb{E}[Y]}{2} \right) ,
\label{eq:av_prop}
\end{equation}
where $\mathbb{E}[\cdot]$ is the expectation operator.

\section{Network performance analysis}
In this section, we evaluate the performance of the shared access network in terms of the throughput of the secondary nodes and the AoI of the primary node.
In particular, we first derive the expressions for the throughput and the time average AoI of the network and then cast two optimization problems.
The first, considers the optimal primary and secondary access probabilities, $q_{pr}$ and $q_s$, that minimize the AoI of the primary node under secondary throughput constraints.
The second problem finds the tuple of optimal primary and secondary access probabilities, $q_{pr}$ and $q_s$, that maximizes the secondary throughput under primary AoI constraints.
The impact of the design parameters is studied in the next section.

\subsection{Service rate of the primary node and secondary throughput}

The service rate of the primary node is given by the following expression
\begin{equation}
\mu = \sum_{k=0}^{N} \binom{N}{k} q_{pr} (1-q_s)^{N-k} q_s^{k} \: P^{d_p}_{p/k},
\label{eq:mu}
\end{equation} 
where $P^{d_p}_{p/k}$ is the success probability of the primary link given that $k$ secondary nodes are transmitting, obtained from \eqref{eq:success_Pr}.
We note that stability is achieved if and only if $\lambda < \mu$. 
The stationary distribution of the primary queue is given in the next subsection.


To derive the secondary throughput we denote by $P^{d_s}_{s,0/k}$ and $P^{d_s}_{s,1/k}$ the success probabilities of a secondary link given that $k$ nodes are transmitting, when the primary node remains silent or transmits respectively.
These success probabilities are obtained from \eqref{eq:success_Pr} and \eqref{eq:symmetric_success_Pr}, where a symmetric $N$-nodes secondary network is assumed.
Let $Q$ denote the queue size of the primary node.
Denote $\mu_s$ the throughput of a secondary node. 
Then, we have
\begin{equation}
\mu_s = \mu_{s,0} \mathbb{P}[Q=0] + (1-q_{pr}) \mu_{s,0} \mathbb{P}[Q\neq 0] + q_{pr} \mu_{s,1} \mathbb{P}[Q\neq0], 
\label{eq:mu_s}
\end{equation} 
where $\mathbb{P}[Q=0]=1-\lambda/\mu$ and $\mathbb{P}[Q\neq0]=1-\mathbb{P}[Q=0]$.
The terms $\mu_{s,0}$ and $\mu_{s,1}$ denote the throughput of a secondary node when the primary node remains silent or transmits respectively.
These are given by 
\begin{equation}
\mu_{s,0} = \sum_{k=0}^{N-1} \binom{N-1}{k} (1-q_s)^{N-k-1} q_s^{k+1} \: P^{d_s}_{s,0/k+1}, 
\end{equation} 
\begin{equation}
\mu_{s,1} = \sum_{k=0}^{N-1} \binom{N-1}{k} (1-q_s)^{N-k-1} q_s^{k+1} \: P^{d_s}_{s,1/k+1}.
\end{equation} 

\subsection{Age of information analysis for the primary node}
\label{sec:analysis}
Next, we derive the average AoI \eqref{eq:av_prop} of the primary node at the destination.
The interarrival times $Y_j$ are i.i.d. sequences that follow a geometric distribution therefore we know that 
\begin{align}
\mathbb{E}[Y_j] & =\frac{1}{\lambda},  \qquad
\mathbb{E}[Y_j^2]  = \frac{2-\lambda}{\lambda^2}.
\label{eq:E_Y_Y2}
\end{align}
Then, the only unknown term for the calculation of the average age is the expectation $\mathbb{E}[YT]$.
The system time of update $j$ is $T_j=W_j+S_j$, where $W_j$ and $S_j$ are the waiting time and service time of update $j$, respectively. 
Since, the service time, $S_j$, is independent of the interarrival time, $Y_j$, we can write
\begin{equation}
\mathbb{E}[Y_j T_j] = \mathbb{E}[Y_j(W_j+S_j)] = \mathbb{E}[Y_j W_j] + \mathbb{E}[Y_j]\mathbb{E}[S_j],
\label{eq:expTiYi}
\end{equation}
where $\mathbb{E}[S_j] = 1/ \mu$.
Moreover, we can express the waiting time of update $j$ as the remaining system time of the previous update minus the elapsed time between the generation of updates $(j-1)$ and $j$, i.e.,
\begin{equation}
W_j = (T_{j-1} - Y_j)^+.
\label{eq:W_i}
\end{equation}
Note that if the queue is empty then $W_j = 0$.
Also note that when the system reaches steady state the system times are stochastically identical, i.e., $T =^{st} T_{j-1} =^{st} T_j$.

In addition, the queue of the primary node can be described through a discrete-time Markov chain, where each state represents the number of packets in the queue. 
Then, it can be shown that the steady state probabilities of the primary node
are
\begin{equation}
\pi_n = \rho^{n-1} \pi_1, \quad n \geq 1,   \quad \text{and} \quad \pi_0 = \frac{\mu(1-\lambda)}{\lambda} \pi_1,
\label{eq:pi_i}
\end{equation}
where $\rho = \frac{\lambda(1-\mu)}{\mu(1-\lambda)}$ and $\pi_1 = \frac{\lambda(1-\rho)}{\mu}$.

To derive the density of the system time $T$, we use the fact that the geometric sum of geometric random variables is geometrically distributed, according to the convolution property of their generating functions \cite{Nelson2013probability}.
Let $S_j, j = 1,2, .. $ be independent and identically distributed geometric random variables with parameter $\mu$.
If an arriving packet sees $n$ packets in the system, then, the system time of that packet, using the memoryless property, can be written as the random sum $T = S_1 + \dots + S_{n}$.
Since $n$ is geometric with parameter $1-\rho$ it follows that $T$ is geometric with parameter $\mu (1-\rho)$. 
This implies that the system time density is given by 
\begin{equation}
f_T(t)= \mu (1-\rho) (1-\mu+\mu \rho)^{t-1}.
\label{eq:f_T}
\end{equation}	

Now we are able to compute the conditional expectation of the waiting time $W_j$ given $Y_j=y$ as
\begin{align}
\mathbb{E}[&W_j|Y_j=y] = \mathbb{E}[(T_{j-1}-y)^+|Y_j=y] = \mathbb{E}[(T-y)^+] = \notag \\
& = \sum_{t=y}^\infty (t-y) f_T(t) =
  \frac{(1-\mu+\mu\rho)^y}{\mu (1-\rho)}.
\label{eq:Wi_cond_Yi}
\end{align}
Then, the expectation $\mathbb{E}[W_j Y_j]$ is obtained as
\begin{align}
\mathbb{E}[W_j Y_j] &=  \sum_{y=0}^\infty y\: \mathbb{E}[W_j | Y_j=y] \: f_{Y_j}(y) = \notag \\ & = \frac{ \lambda (1-\mu+\mu \rho)}{\mu (1-\rho) (\lambda+\mu-\lambda \mu-\mu \rho+\lambda \mu \rho)^2}.
\label{eq:expWiYi}
\end{align}
Substituting $\rho = \frac{\lambda(1-\mu)}{\mu(1-\lambda)}$ to \eqref{eq:expWiYi} and after some algebra we obtain
\begin{equation}
\mathbb{E}[W_j Y_j] = \frac{ \lambda (1-\mu)}{(\mu- \lambda) \mu^2}.
\label{eq:expWiYi2}
\end{equation}
From \eqref{eq:E_Y_Y2}, \eqref{eq:expTiYi}, and \eqref{eq:av_prop}, the average AoI of the primary node is obtained as
\begin{equation}
\Delta  = \frac{1}{\lambda}+\frac{1-\lambda}{ \mu-\lambda}-\frac{\lambda}{\mu^2}+\frac{\lambda}{\mu}.
\label{eq:Delta}
\end{equation}

In order to find the optimal value of $\lambda$ that minimizes the average AoI we proceed as follows.
We differentiate \eqref{eq:Delta} with respect to $\lambda$ to obtain $\frac{\partial \Delta}{\partial \lambda}$. 
By setting $\frac{\partial \Delta}{\partial \lambda}=0$ we can obtain the value of $\lambda$ that minimizes the AoI and satisfies the equation $\lambda^4 (\mu-1) - 2 \lambda^3 (\mu-1) \mu- \lambda^2 \mu^2 +2 \lambda \mu^3 -\mu^4 = 0$.
Taking the second derivative $\frac{\partial^2 \Delta}{\partial \lambda^2}$ it can be easily shown that $\Delta$ is a convex function of $\lambda$ for a given service rate $\mu$, if $\lambda<\mu$ is not violated. 
The proof is omitted due to space limitations.

\section{Optimization formulation}
\subsection{Minimum age subject to throughput  requirements}
To achieve the optimal performance of the network, we present the optimization problem of minimizing the average AoI of the primary node, while guaranteeing a certain throughput for the secondary nodes.
This can be done by adjusting properly the access probabilities $q_s$ and $q_{pr}$.
We define the aggregate secondary throughput as $\mu_{\text{total}} = N\:\mu_s$.
Then, the general optimization formulation is given by
\begin{align}
\min_{q_s,q_{pr}} \:\:\:& \Delta \label{eq:age_obj}\\
\text{subject to} \:\:\:& \mu_{\text{total}} \geq \mu_{\text{min}}, \label{eq:thr_constr} \\
\:\:\:& \Delta \geq 0, \\
\:\:\:& q_s, q_{pr} \in [0,1], 
\end{align}
where $\mu_{\text{min}}$ is the minimum required aggregate throughput in packets/slot and $\Delta$ is obtained by substituting \eqref{eq:mu} to \eqref{eq:Delta} in order to express it in terms of the variables $q_s$ and $q_{pr}$. 

\subsection{Maximum throughput subject to an age constraint}
Next, we present the optimization problem of maximizing the aggregate throughput of the secondary nodes $\mu_{\text{total}}$, while restricting the average AoI of the primary node to remain below a certain threshold.
The general optimization formulation is given by
\begin{align}
\max_{q_s,q_{pr}} \:\:\:& \mu_{\text{total}} \label{eq:thr_objective}\\
\text{subject to} \:\:\:& 0 \leq \Delta \leq \Delta_{\text{max}}, \label{eq:age_constr} \\
\:\:\:& q_s, q_{pr} \in [0,1], 
\end{align}
where $ \Delta_{\text{max}}$ is the maximum average staleness the primary node can tolerate.

These optimization problems are difficult to solve analytically due to the complexity of the expressions related to the secondary throughput $\mu_s$ and the average age $\Delta$.
Therefore, in Section~\ref{sec:results} we present a numerical evaluation of the problems.

\section{Numerical Results}
\label{sec:results}

We consider a cognitive shared access network where the primary node is located at distance $r(p,d)=150$ m from the destinations $d_{p}$, $d_s$.
The secondary nodes are located at isotropic directions around their associated destinations with fixed distance $r(s,d) =40$ m.
The primary and secondary transmission powers are $P_{tx}(p,d) = 10$ mW and $P_{tx}(s,d) = 0.1$ mW, respectively.
The receiver noise power is  $\eta_d = -121$ dBm.
The path loss exponent is $\alpha = 4$.

In order to obtain a better understanding of the optimization results and the objectives considered, we illustrate the two performance metrics of interest as functions of different system parameters.

\begin{figure}[ht]\centering
	\centering
	\includegraphics[draft=false,scale=.5]{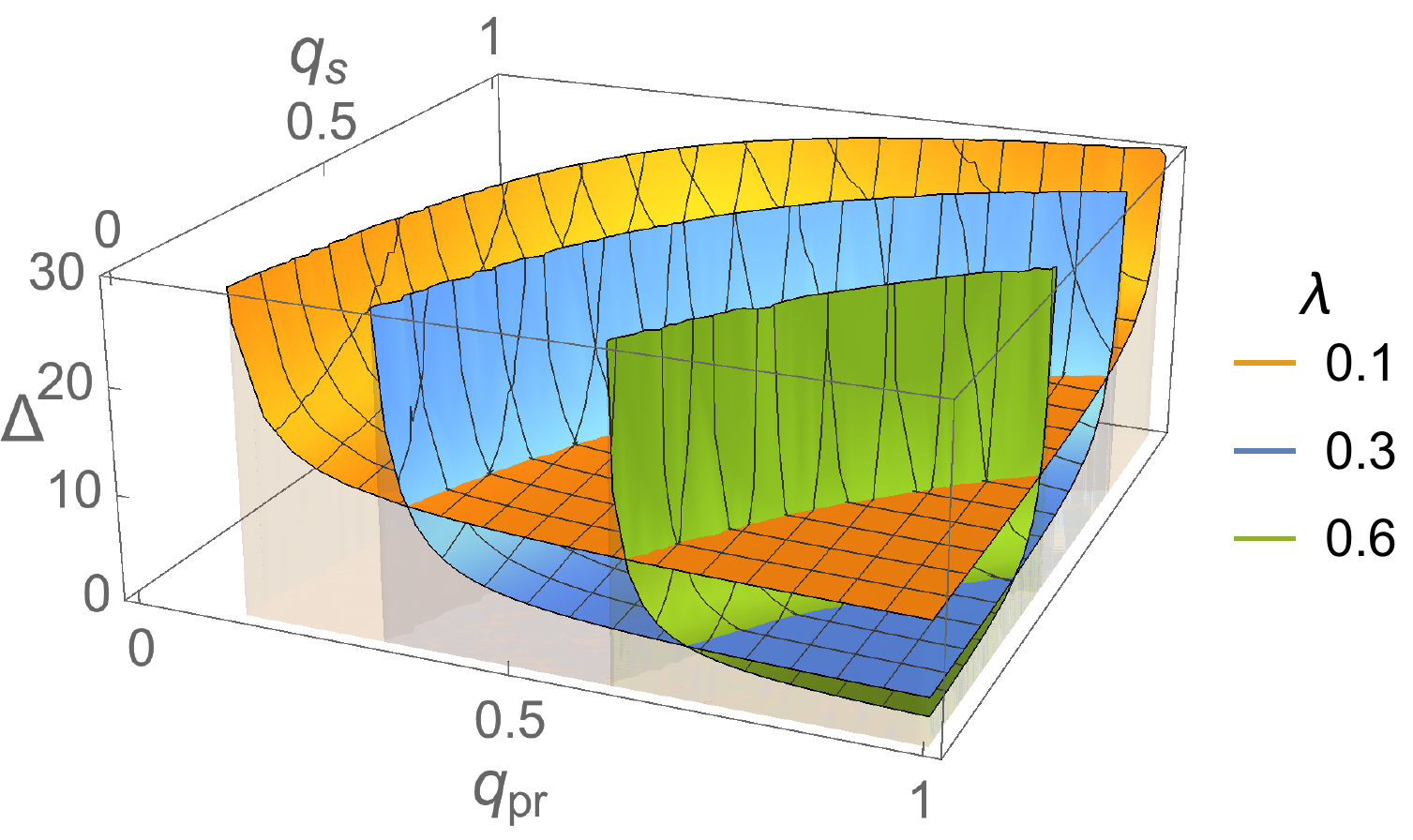}
	\caption{Average age $\Delta$ as a function of the secondary access probability $q_s$ and the primary access probability $q_{pr}$ for various values of the arrival rate $\lambda$.}
	\label{fig:Delta_vs_q_s_q_p_3D}
\end{figure}

\begin{figure}[ht]\centering
	\centering
	\includegraphics[draft=false,scale=.5]{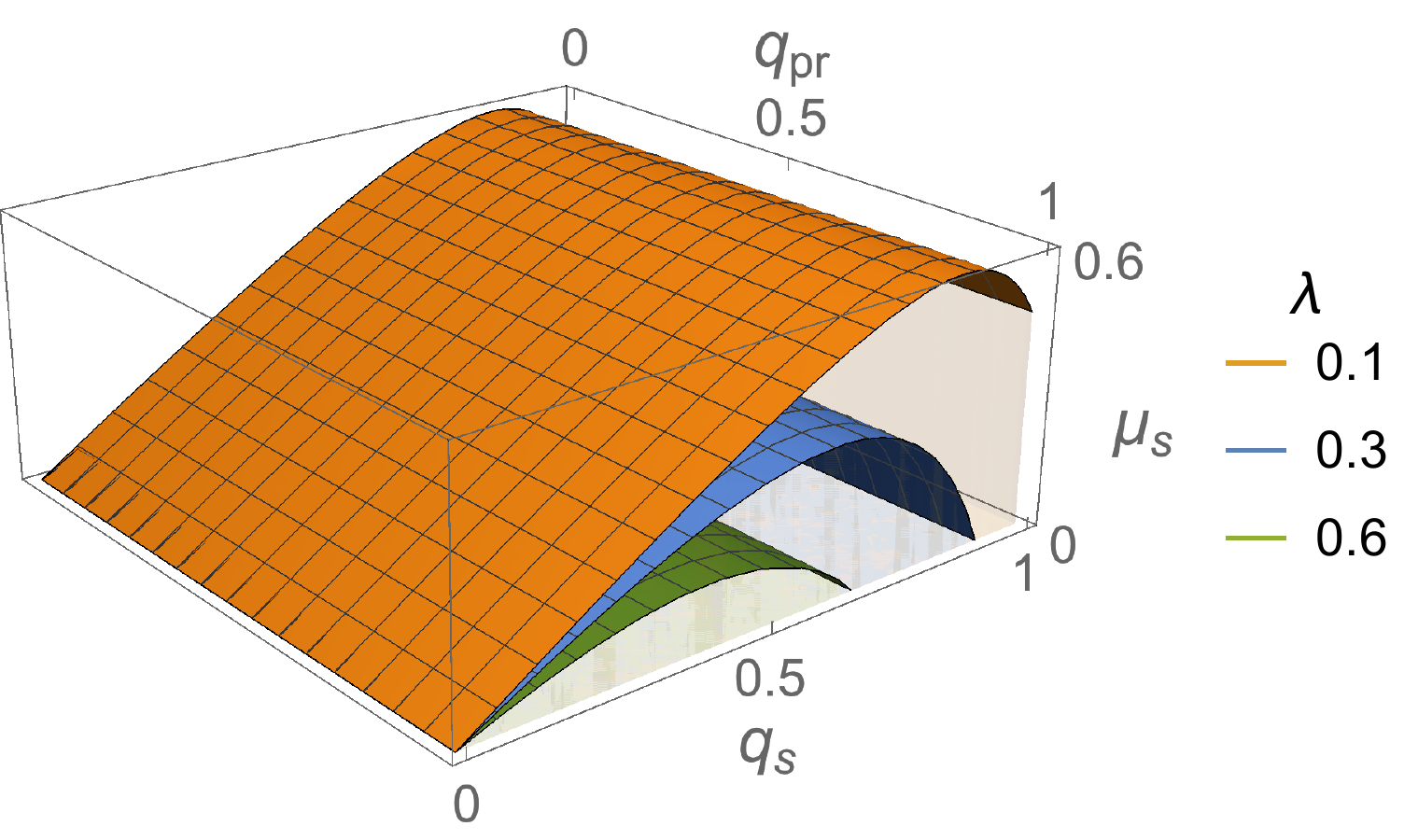}
	\caption{Secondary throughput $\mu_s$ as a function of the secondary access probability $q_s$ and the primary access probability $q_{pr}$ for various values of the arrival rate $\lambda$.}
	\label{fig:Ts_vs_q_p_and_q_s_3D}
\end{figure}

\begin{figure}[ht]\centering
	\centering
	\includegraphics[draft=false,scale=.5]{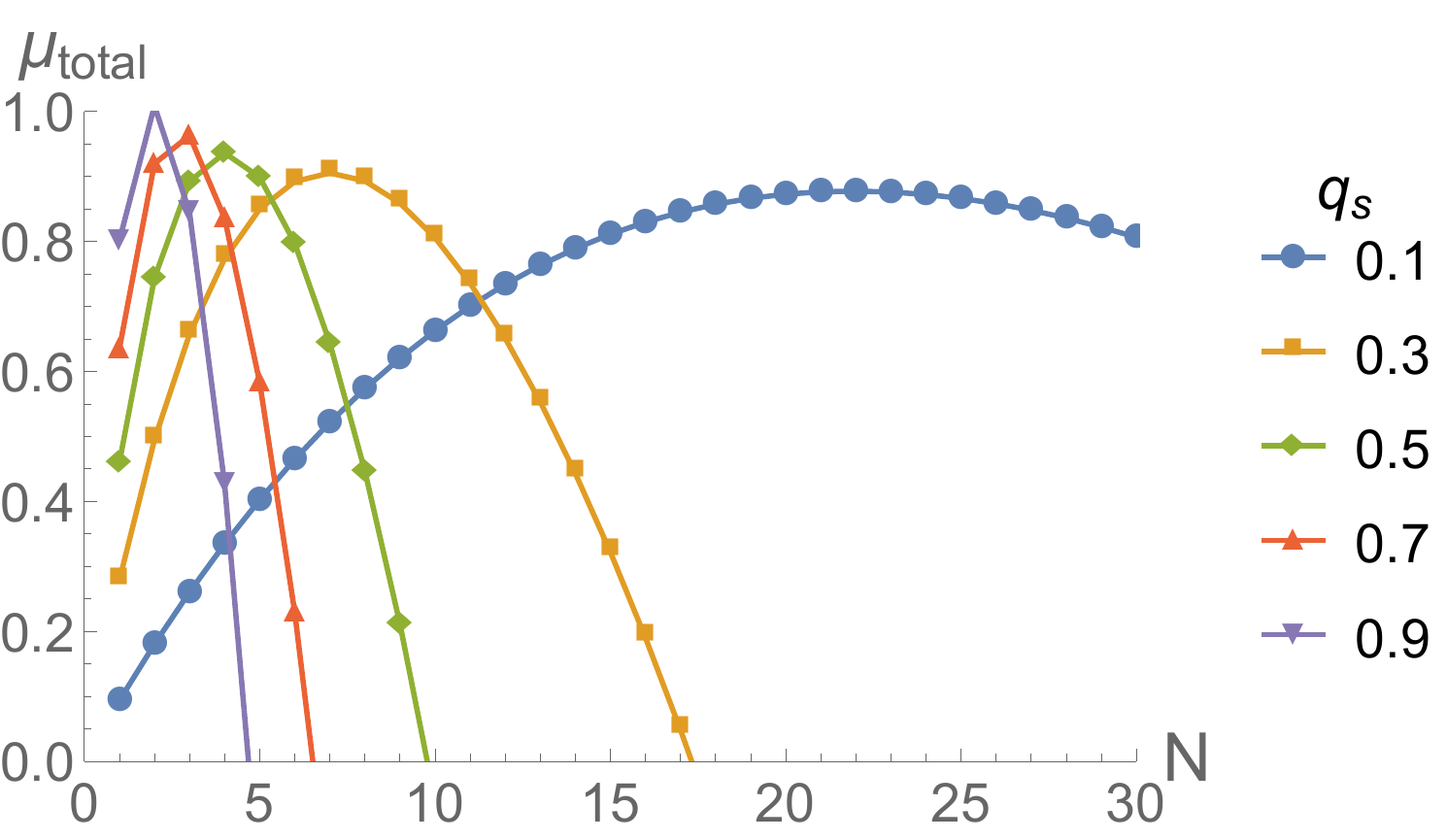}
	\caption{Secondary aggregate throughput $\mu_{\text{total}}$ as a function of the number of secondary nodes $N$ for various values of the secondary access probability $q_s$.}
	\label{fig:Ttotal_vs_N_and_q_s_2D}
\end{figure}

In Fig.~\ref{fig:Delta_vs_q_s_q_p_3D}, the average AoI of equation \eqref{eq:Delta} is depicted as a function of the primary access probability $q_{pr}$ and the secondary access probability $q_s$ for three values of the arrival rate $\lambda$.
The $SINR_i$ target $\gamma_d$ is 5 dB and $N=1$.
The age $\Delta$ is a monotonic function of $q_{pr}$ and $q_s$. 
The conditions that guarantee the stability of the primary queue also determine the feasibility region of age. We observe that higher rate $\lambda$ results in smaller AoI and smaller feasible region.

In Fig.~\ref{fig:Ts_vs_q_p_and_q_s_3D}, the secondary throughput of equation \eqref{eq:mu_s} is depicted as a function of the primary access probability $q_{pr}$ and the secondary access probability $q_s$ for three values of the arrival rate $\lambda$.
The $SINR_i$ target $\gamma_d$ is 5 dB and $N=1$.
We note that $\mu_s$ is independent of $q_{pr}$ when the queue at the primary node is stable.
In this case, the secondary throughput decreases with the arrival rate $\lambda$.
Moreover, we note that $\mu_s$ is not a monotonic function of $q_s$.
There exists an optimal value of $q_s$ that gives the maximum $\mu_s$ among the feasible choices.

In Fig.~\ref{fig:Ttotal_vs_N_and_q_s_2D}, we show the aggregate throughput $\mu_{\text{total}}$ as a function of the number of secondary nodes $N$ for different access probabilities $q_s$.
The $SINR_i$ target $\gamma_d$ is -3 dB and $\lambda=0.3$.
We observe that there is an optimum number of user $N^*$ that maximizes the aggregate throughput and depends on the access probability $q_s$.
This number can be used as a criterion of the subset of nodes that should be active in the network.

\begin{figure}[h!]\centering
	\centering
\begin{tikzpicture}
\node[anchor=south west,inner sep=0] at (0,0) {\includegraphics[draft=false,scale=.485]{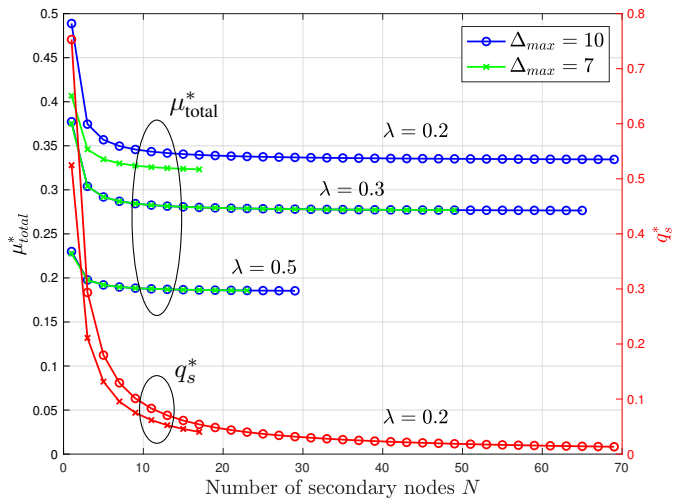}};
\draw[black] (2.0,3.75) ellipse  (0.33cm and 1.3cm);
\draw	(2.5,5.55) node[anchor=north] {$\mu_{\text{total}}^*$};
\draw[black] (2.0,1.2) ellipse  (0.23cm and 0.45cm);
\draw	(2.4,2.0) node[anchor=north] {$q_s^*$};
\end{tikzpicture}
	\caption{Optimal aggregate throughput $\mu_{\text{total}}^*$ as a function of the number of secondary nodes $N$.}
\label{fig:Ttotal_opt_vs_N}
\end{figure}

Furthermore, in Fig.~\ref{fig:Ttotal_opt_vs_N} we solve problem \eqref{eq:thr_objective} and plot the optimal aggregate throughput $\mu_{\text{total}}^*$ and the optimal secondary access probability $q_s^*$ as the number of secondary nodes $N$ in the system increases, for different arrival rates $\lambda$ and timeliness constraints $\Delta_{\text{max}}$.
Making the age requirement more strict decreases the number of secondary nodes that can be served for a given arrival rate at the primary queue.
For fixed $\lambda$, up to a certain threshold of $\Delta_{max}$ the aggregate throughput remains the same.
When $\Delta_{max}$ decreases beyond that threshold $\mu_{\text{total}}^*$ decreases as well.
An interesting trade-off that derives from the characteristics of age is illustrated for $\Delta_{max}=7$.
That is, $\lambda=0.2$ leads to higher aggregate throughput than $\lambda=0.3$ but less served nodes.
This can be explained by Fig.~\ref{fig:Ts_vs_q_p_and_q_s_3D} where higher arrival rate leads to smaller $\mu_s$. 
However, for fixed $\mu$ the age $\Delta$ is a convex function with respect to the arrival rate $\lambda$.
This non-monotonicity differs age from other performance metrics such us delay or throughput.
This means that for fixed $\Delta_{max}$ these is an optimum $\lambda$ that leads to the largest feasible region of $(N,\lambda)$.
Moreover, we plot the optimal secondary access probability $q_s^*$  for $\lambda=0.2$ and note that restricting the AoI constraint leads to smaller values for $q_s^*$.
The optimal primary access probability $q_{pr}^*$ tends to 1 for problem \eqref{eq:thr_objective}.

\begin{figure}[ht]\centering
	\centering
	\begin{tikzpicture}
	\node[anchor=south west,inner sep=0] at (0,0) {\includegraphics[draft=false,scale=.485]{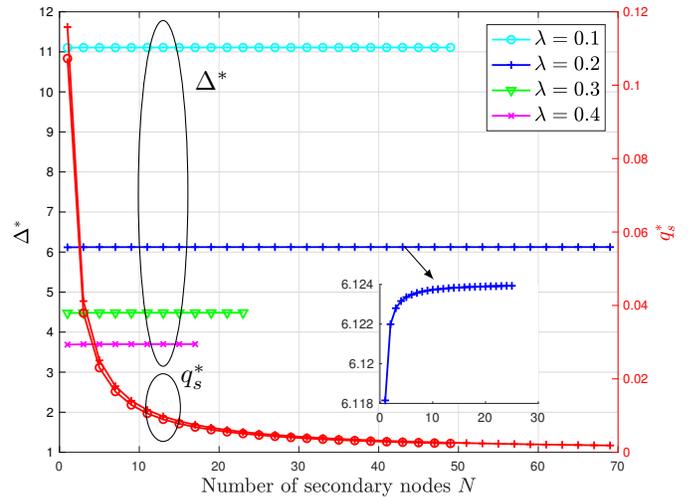}};
	\draw[black] (2.0,4.05) ellipse  (0.33cm and 2.3cm);
	\draw	(2.65,5.8) node[anchor=north] {$\Delta^*$};
	\draw[black] (2.0,1.2) ellipse  (0.23cm and 0.45cm);
	\draw	(2.4,1.9) node[anchor=north] {$q_s^*$};
    \draw[<-] (5.6,2.9) -- (5.2,3.35);        
	\end{tikzpicture}
	\caption{Optimal average age $\Delta^*$ as a function of the number of secondary nodes $N$.}
	\label{fig:Delta_opt_vs_N}
\end{figure}

In Fig.~\ref{fig:Delta_opt_vs_N}, we plot the solution obtained from \eqref{eq:age_obj} as a function of the number of secondary nodes $N$, where the minimum required secondary throughput is $\mu_{min}=0.1$.
As expected, higher $N$ leads to higher average age $\Delta^*$.
On the other hand, for fixed $N$, increasing $\lambda$ results in smaller $\Delta^*$, assuming there is a feasible solution to the problem.
Similarly to problem \eqref{eq:thr_objective}, there is a trade-off between the AoI of the primary node and the subset of active secondary nodes.
Moreover, we plot the optimal secondary access probability $q_s^*$ for $\lambda=0.1$ and $\lambda=0.2$, and note that increasing $\lambda$ leads to higher values for $q_s^*$.
The optimal primary access probability $q_{pr}^*$ is $1$ for problem \eqref{eq:age_obj}.

\setlength\textfloatsep{1.25\baselineskip plus 3pt minus 2pt}

\section{Conclusion}
\label{sec:conclusions}
In this work, we have considered the characterization of the time average age of information of the primary node and the throughput of $N$ secondary nodes in a shared access network with priorities. We derived analytical expressions for both performance metrics and addressed the problem of optimizing our network with respect to them. 
In particular, we formulated the minimum average AoI optimization problem  subject to an aggregate secondary throughput requirement, as well as the maximum aggregate secondary throughput optimization problem subject to a maximum time average staleness constraint.
Our results illustrate the impact of the system parameters on the performance and indicate
that there is an optimum system-dependent arrival rate at the primary node that leads to the largest feasible region of $(N,\lambda)$.
The rate $\lambda$ that maximizes the feasible region $(N,\lambda)$ is not necessarily optimal with respect to the performance objective of the network.

\section*{Acknowledgment}
The research leading to these results has been partially funded by the
European Union's Horizon 2020 research and innovation programme under
the Marie Sklodowska-Curie Grant Agreement No. 642743 (WiVi-2020).

\vspace{+0.06cm}

\bibliography{references}
\bibliographystyle{IEEEtran}

\end{document}